\documentclass[%
 preprint,
superscriptaddress,
 amsmath,amssymb,
 aps,
 prb,
floatfix,
]{revtex4-2}

\usepackage{physics}
\usepackage{booktabs}
\usepackage{braket}
\usepackage{qcircuit}
\usepackage{graphicx}
\usepackage{dcolumn}
\usepackage{bm}
\usepackage[hidelinks]{hyperref}
\usepackage{tensor}
\usepackage{xcolor}
\usepackage{threeparttable}
\usepackage{chemformula}

\begin{document}


\title{Autonomous Quantum Simulation through Large Language Model Agents}
\newcommand{\CUHK}{School of Science and Engineering, The Chinese University of Hong Kong, Shenzhen, Guangdong 518172, China}
\newcommand{\THU}{Department of Chemistry, MOE Key Laboratory for Organic OptoElectronics and Molecular Engineering, Tsinghua University, Beijing 100084, China}
\newcommand{\IOP}{Institute of Physics, Chinese Academy of Sciences, Beijing 100190, China}
\newcommand{\Aggre}{Guangdong Basic Research Center of Excellence for Aggregate Science, School of Science and Engineering, The Chinese University of Hong Kong, Shenzhen, Shenzhen, Guangdong, 518172, P.R. China}

\author{Weitang Li}
\email{liwt31@gmail.com}
\affiliation{\Aggre}

\author{Jiajun Ren}
\affiliation{MOE Key Laboratory of Theoretical and Computational
Photochemistry, College of Chemistry, Beijing Normal University,
Beijing, 100875, P. R. China}

\author{Lixue Cheng}
\affiliation{Department of Chemistry, The Hong Kong University of Science and Technology, Hong Kong, P. R. China}

\author{Cunxi Gong}
\affiliation{\Aggre}

\date{\today}

\begin{abstract}
We demonstrate that large language model (LLM) agents can autonomously perform tensor network simulations of quantum many-body systems, achieving approximately 90\% success rate across representative benchmark tasks. Tensor network methods are powerful tools for quantum simulation, but their effective use requires expertise typically acquired through years of graduate training. By combining in-context learning with curated documentation and multi-agent decomposition, we create autonomous AI agents that can be trained in specialized computational domains within minutes. We benchmark three configurations (baseline, single-agent with in-context learning, and multi-agent with in-context learning) on problems spanning quantum phase transitions, open quantum system dynamics, and photochemical reactions. Systematic evaluation using DeepSeek-V3.2, Gemini 2.5 Pro, and Claude Opus 4.5 demonstrates that both in-context learning and multi-agent architecture are essential. Analysis of failure modes reveals characteristic patterns across models, with the multi-agent configuration substantially reducing implementation errors and hallucinations compared to simpler architectures.
\end{abstract}

\maketitle

\section{Introduction}

Large language models (LLMs) have emerged as transformative tools for scientific research~\cite{jablonka202314, minaee2024large, wang2024survey, xi2025rise, ramos2025review}. Models such as GPT-4~\cite{achiam2023gpt}, Claude, Gemini~\cite{team2023gemini}, and the open-weight DeepSeek model~\cite{guo2025deepseek} demonstrate remarkable capabilities across diverse scientific tasks~\cite{ai4science2023impact, jablonka2024leveraging}. When deployed as autonomous agents with external tools, memory, and planning abilities~\cite{yao2022react}, these models become substantially more powerful. Tool augmentation through approaches like Toolformer~\cite{schick2023toolformer} and retrieval-augmented generation~\cite{gao2023retrieval} equips LLMs with external capabilities while reducing hallucinations. Building on decades of work toward self-driving laboratories~\cite{steiner2019organic, abolhasani2023rise, song2025multiagent}, LLM-driven agents have achieved notable successes across scientific domains. Examples include ChemCrow, which outperforms GPT-4 alone on chemical research tasks~\cite{m2024augmenting}; Coscientist, which autonomously executes chemical syntheses~\cite{boiko2023autonomous, huang2025natural}; and El Agente Q, which achieves 88\% success on quantum chemistry exercises~\cite{zou2025agente}. Other agents have advanced quantum chemistry accessibility~\cite{hu2025aitomia, gadde2025chatbot}, organic semiconductor optimization~\cite{zhang2024large}, single-cell transcriptomics~\cite{xie2025cassia}, autonomous microscopy~\cite{mandal2025evaluating}, and nanobody design~\cite{swanson2025virtual}.

Tensor network methods represent another powerful paradigm in computational science~\cite{orus2019tensor, banuls2023tensor}. By exploiting entanglement structure, tensor networks overcome the exponential barrier that prohibits direct classical simulation of quantum many-body systems. Over three decades, algorithmic development from White's density matrix renormalization group (DMRG)~\cite{white1992density, white1993density} and matrix product states (MPS)~\cite{schollwock2011density} to tree tensor networks~\cite{shi2006classical, vidal2007entanglement, cirac2021matrix} and time-evolution methods like the time-dependent variational principle (TDVP)~\cite{haegeman2011time, haegeman2016unifying, lubich2015tt, paeckel2019time} has produced a rich ecosystem of methods tailored to quantum chemistry~\cite{li2019electronic, sharma2020magnetic, larsson2022chromium}, condensed matter physics~\cite{liao2017gapless, fowler2022efficient}, and quantum computing~\cite{huang2021efficient, berezutskii2025tensor}. This diversity is reflected in more than 30 actively maintained software packages~\cite{sehlstedt2025software}. The proliferation of specialized tools underscores both the versatility and the complexity of tensor network methods, as effective use requires not only selecting an appropriate package but also mastering its particular conventions for network structures~\cite{nakatani2013efficient, rams2020breaking}, bond dimension control~\cite{tagliacozzo2009simulation, eisert2010colloquium}, symmetry implementations~\cite{chan2004algorithm, weichselbaum2012non, li2023tangent}, and observable computation.

Despite the individual strengths of LLMs and tensor networks, applying LLMs to automate tensor network simulations presents substantial challenges. Tensor network simulation requires simultaneous reasoning across three closely integrated layers: the underlying physics, code implementation, and numerical data. This physics-code-data coupling creates three difficulties. First, tensor network methods receive sparse coverage in LLM training data, causing severe hallucinations when models attempt simulations from incomplete knowledge. Second, simulations produce dense numerical outputs requiring precise quantitative analysis that LLMs cannot reliably perform through direct reasoning. Third, the tight coupling makes it difficult for a single agent to validate its own results, as failures may involve physics, code, or data artifacts simultaneously.

We address these challenges through in-context learning~\cite{dong2024survey} with curated documentation and multi-agent decomposition for specialized validation. We systematically compare three configurations: a baseline single agent without documentation, a single agent with in-context learning, and a multi-agent architecture combining both approaches. Our multi-agent system interfaces with Renormalizer~\cite{renormalizer, ren2018time}, a general tensor network package with a focus on electron-phonon dynamics~\cite{li2020finite, li2021general, wang2023minimizing, sheng2024td}. Benchmarking on quantum phase transitions in the two-dimensional Ising model, spin dynamics in the sub-Ohmic spin-boson model, and retinal photoisomerization using DeepSeek-V3.2, Gemini 2.5 Pro, and Claude Opus 4.5 reveals that in-context learning is critical for success and that multi-agent decomposition further improves reliability. Different models exhibit characteristic failure modes.

\section{Results}

\subsection{Agent Architecture}

Our approach combines two key strategies to address the challenges of autonomous tensor network simulation: in-context learning and multi-agent decomposition.
The first strategy, in-context learning, addresses the sparse coverage of tensor network methods in LLM training data. Rather than relying on retrieval-augmented generation or expensive fine-tuning, we embed carefully curated Renormalizer documentation directly in the system prompt, totaling approximately 43k tokens. This documentation comprises Jupyter notebook tutorials (22k tokens), Python script examples (12k tokens), and refactored source code snippets (9k tokens). Embedding comprehensive documentation ensures that agents have instant access to all relevant information without relying on potentially imperfect retrieval queries. The complete Renormalizer source code would require approximately 0.8 million tokens, nearly 20 times the size of the curated documentation, and raw retrieval from source code often returns fragmented results or those lacking context that confuse the agent rather than help it. When debugging is required, agents can still access the full source code through file reading tools.
While our implementation targets Renormalizer, the approach generalizes to other tensor network packages. The key requirement is comprehensive documentation that can fit within the context window. We expect similar agent systems can be developed for packages such as ITensor~\cite{fishman2022itensor}, TeNPy~\cite{hauschild2018efficient}, and Block2~\cite{zhai2023block2} with comparable effort.

The second strategy, multi-agent decomposition, addresses the tight coupling between physics, code, and data that makes validation difficult for a single agent. Scientific workflows naturally decompose into specialized subtasks, from high-level research planning to code implementation and result analysis. A single agent attempting to handle all aspects must maintain lengthy context and switch frequently between distinct reasoning modes. Our multi-agent architecture instead assigns each subtask to a dedicated agent with its own isolated context, a design pattern known as context quarantine. This isolation enables clear separation of concerns and more focused prompts, preventing irrelevant information from one phase of the workflow from interfering with reasoning in another.

\begin{figure}[htbp]
\centering
\includegraphics[width=\textwidth]{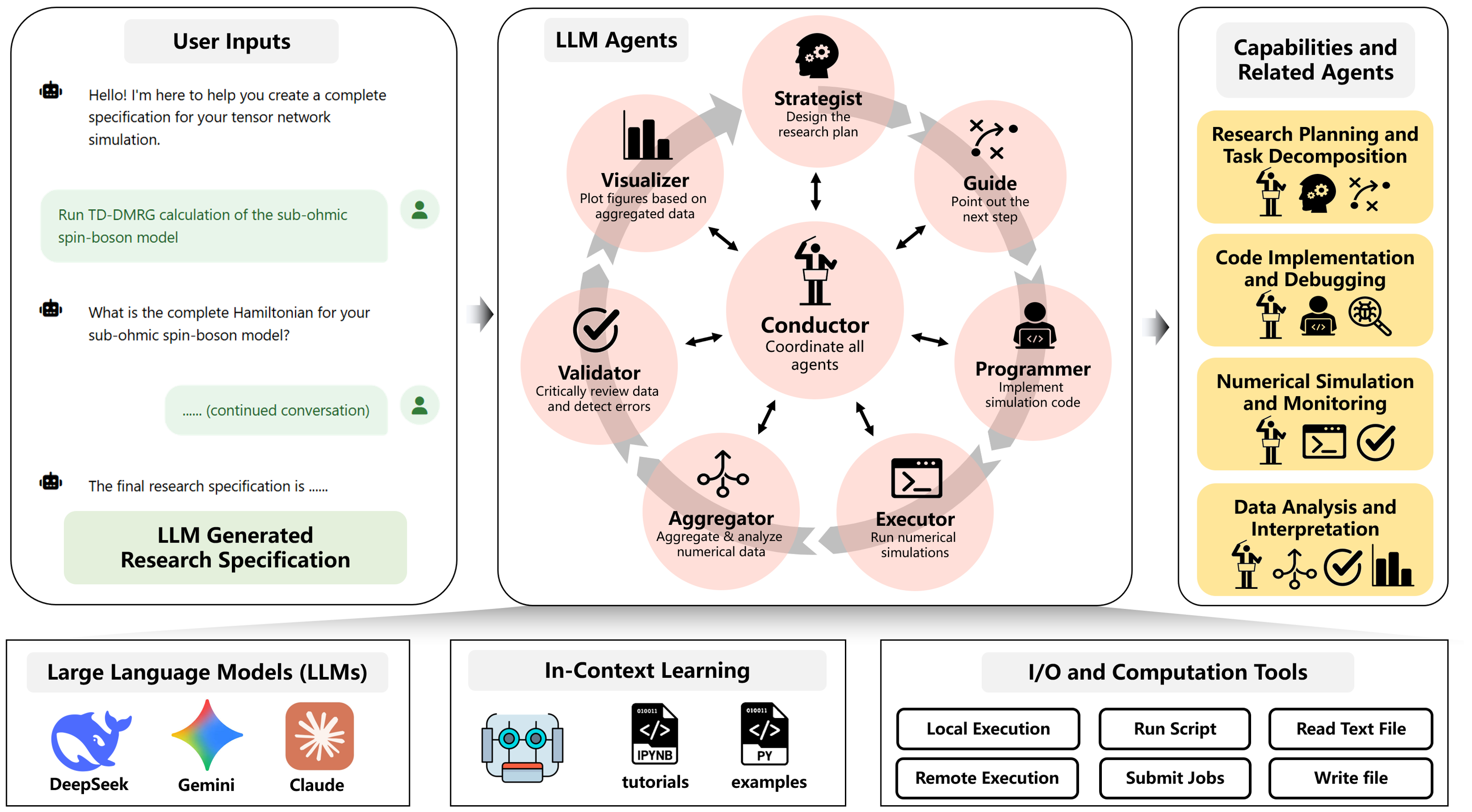}
\caption{\textbf{Multi-agent architecture for autonomous tensor network simulation.} The \textbf{Conductor} agent orchestrates seven specialized agents: \textbf{Strategist}, \textbf{Guide}, \textbf{Programmer}, \textbf{Executor}, \textbf{Aggregator}, \textbf{Validator}, and \textbf{Visualizer}. Agents communicate through structured messages and share access to tools such as file operations and code execution. The user interacts through natural language problem specifications and receives publication-ready outputs.}
\label{fig:architecture}
\end{figure}

As shown in Fig.~\ref{fig:architecture}, our system comprises a central \textbf{Conductor} that coordinates seven specialized agents. The \textbf{Strategist} designs the overall research plan by decomposing complex scientific questions into manageable computational tasks. The \textbf{Guide} monitors progress through the simulation workflow and decides on the next step. The \textbf{Programmer} implements simulation code based on the research specification, with access to the embedded Renormalizer documentation. The \textbf{Executor} runs numerical simulations, monitors job progress, and handles minor debugging. The \textbf{Aggregator} collects and organizes numerical data from completed simulations, then performs data processing and analysis. The \textbf{Validator} critically reviews simulation results, checking for numerical artifacts, convergence issues, and physical inconsistencies. The \textbf{Visualizer} generates publication-quality figures from aggregated data. In the following sections, we include figures generated by the agent without manual modification or post-processing.

The \textbf{Conductor} orchestrates these agents iteratively, routing tasks to appropriate specialists and maintaining coherent progress toward the research objective. When an agent encounters difficulties, the \textbf{Conductor} routes the problem back for revision with relevant error information. For example, when the \textbf{Executor} encounters a bug beyond its capability, the \textbf{Conductor} reports it to the \textbf{Programmer} for fixing. Similarly, when the \textbf{Validator} raises concerns about numerical convergence, the \textbf{Conductor} directs the \textbf{Executor} to perform additional calculations with refined parameters.

The workflow begins with an interactive dialogue between the user and the agent to establish a complete research specification. Through this conversation, the user provides a high-level description of the research goal, and the agent asks clarifying questions about physical parameters, numerical settings, and computational constraints. The resulting specification captures all details needed for simulation and is then fed into the multi-agent system, which autonomously executes the simulation workflow. A detailed example of this process is presented in the following sections as well as in the Supporting Information.

We compare three agent configurations throughout this work.
The baseline configuration uses a single agent without in-context learning, resembling general-purpose coding agents. When attempting simulations, the baseline agent must search and grep through the Renormalizer source code to learn usage patterns, a process prone to retrieving incomplete or misleading information. The single-agent configuration adds curated documentation to the system prompt, providing comprehensive examples and API references. The multi-agent configuration combines in-context learning with the architecture shown in Fig.~\ref{fig:architecture}.
In this work, we evaluate three state-of-the-art LLMs as backend engines, including DeepSeek-V3.2, Gemini 2.5 Pro, and Claude Opus 4.5. All three models achieve high success rates on the benchmark tasks presented below. The architecture is compatible with other LLMs, and we expect comparable frontier models to perform similarly.

\subsection{Benchmark Tasks}

We evaluate our agent on three benchmark tasks spanning different areas of quantum many-body physics, with detailed physics background provided in the Methods section. The two-dimensional transverse-field Ising model represents a canonical ground-state problem with a well-established critical point ($h_c/J \approx 3.04$ in the thermodynamic limit). We study an $8 \times 8$ lattice to ensure calculations complete within a reasonable time frame. This task tests whether agents can correctly implement DMRG calculations and interpret the resulting phase transition. The well-established critical value, however, creates a risk of hallucinations, as LLMs may output expected results even when simulations fail. The sub-Ohmic spin-boson model involves real-time quantum dynamics and phase classification. Unlike the Ising model, the coherent-incoherent phase boundary lacks a definitive literature value despite decades of numerical investigation, making validation more subtle. The fine phase boundary mapping presented in this work constitutes new scientific results rather than reproduction of existing literature. This task tests whether agents can correctly implement time evolution using TDVP, perform convergence analysis, and classify dynamical behavior from numerical trajectories. The retinal photoisomerization model demands implementation of specialized basis sets (exponential DVR) not available in Renormalizer. It also requires careful handling of unit and symbol conventions that differ between the literature and the software. This task tests the agent's ability to translate complex physical specifications into working code, integrating information from multiple sources and understanding implicit conventions in scientific notation. For an expert already familiar with tensor network theory and proficient with a tensor network package, we estimate these tasks would require one day to one week to complete, depending on prior familiarity with the specific physical models.

\begin{figure}[htbp]
\centering
\includegraphics[width=\textwidth]{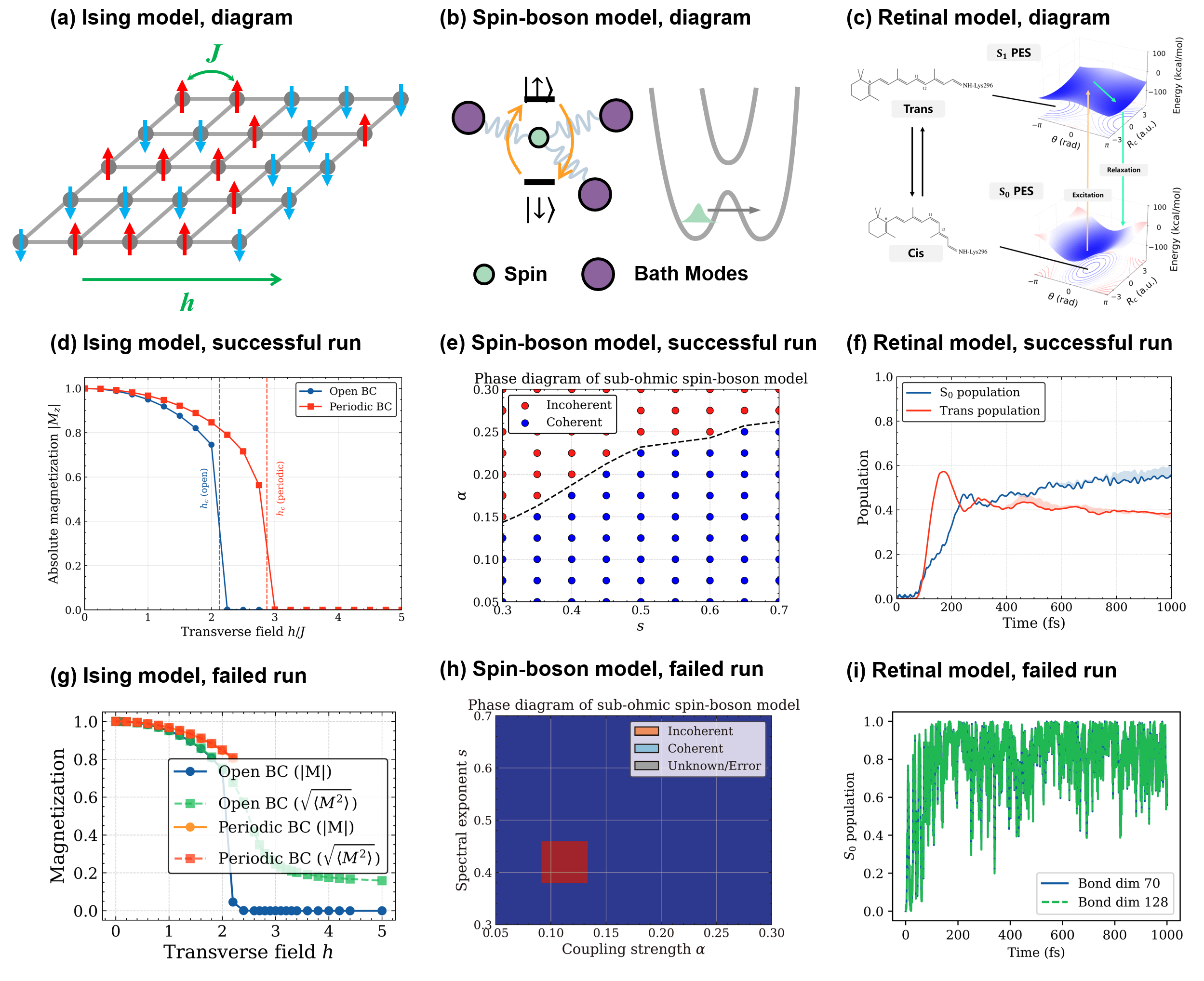}
\caption{\textbf{Model schematics and representative simulation results.} \textbf{(a--c)} Schematic diagrams of the three benchmark models: (a) 2D transverse-field Ising model with ferromagnetic coupling $J$ and transverse field $h$; (b) spin-boson model showing the two-level system coupled to bath modes and the diabatic potential energy surfaces; (c) retinal photoisomerization model with torsional coordinate connecting cis and trans configurations. \textbf{(d--f)} Successful runs from the multi-agent architecture showing correct phase transition behavior, accurate phase classification, and converged population dynamics. \textbf{(g--i)} Failed runs from the baseline configuration illustrating characteristic failure modes, which are missing data and overlapping curves, incomplete phase diagrams dominated by errors, and unphysical oscillatory dynamics.}
\label{fig:cases}
\end{figure}

Figure~\ref{fig:cases} presents model schematics alongside representative successful and failed runs for each task. The top row illustrates the three physical models. Figure~\ref{fig:cases}(a) shows the 2D transverse-field Ising model, where spins on a square lattice interact ferromagnetically with coupling $J$, while a transverse field $h$ induces quantum fluctuations. Figure~\ref{fig:cases}(b) depicts the spin-boson model, where a two-level system with states $|\!\uparrow\rangle$ and $|\!\downarrow\rangle$ couples to a bath of harmonic oscillators, with the right panel showing the diabatic potential energy surfaces as a function of spin polarization $\sigma_z$. Figure~\ref{fig:cases}(c) illustrates the retinal photoisomerization model, where the torsional coordinate $\theta$ connects cis and trans configurations, and the 3D surface shows the coupled S$_0$ and S$_1$ potential energy surfaces that govern the ultrafast isomerization dynamics.

The middle row (d--f) shows successful runs produced by the multi-agent architecture. All figures are generated by the agent without manual modification or post-processing. Figure~\ref{fig:cases}(d) shows absolute magnetization versus transverse field for the Ising model. The figure correctly captures that periodic boundary conditions stabilize the ferromagnetic phase to higher field values due to additional bonds at the boundaries. The user's initial prompt was deliberately brief: ``Run DMRG simulation of 2d ising model and see how $|M_z|$ changes with $h$. compare open and periodic boundary conditions.'' From this minimal input, the agent initiated a clarification process to fill in the technical details required for a rigorous simulation. After receiving all necessary information from the user, the agent compiled a comprehensive research specification and passed it to the \textbf{Conductor} for the simulation workflow. Details of this dialogue are provided in the Supporting Information.
Figure~\ref{fig:cases}(e) shows a successful phase diagram for the spin-boson model. The critical coupling strength $\alpha_c$ for the coherent-to-incoherent transition increases monotonically with the spectral exponent $s$, consistent with the physical expectation that sub-Ohmic baths with smaller $s$ induce stronger dissipation. The agent autonomously performed parameter searches to determine appropriate bond dimensions and the number of phonon modes, calculated the dynamics across the parameter grid, and classified each point as coherent or incoherent based on the specified criteria.
Figure~\ref{fig:cases}(f) shows successful retinal photoisomerization dynamics. The S$_0$ population and trans population evolution are consistent with reference calculations from the literature~\cite{liu2024benchmarking}, demonstrating that the agent correctly implemented the exponential DVR basis and handled the coordinate conventions properly.

The bottom row (g--i) shows failed runs from the baseline configuration, illustrating characteristic failure modes. Figure~\ref{fig:cases}(g) exhibits hallucinations: the agent prematurely concluded that the simulation was finished, resulting in missing periodic boundary condition data. The agent also computed $\sqrt{\langle M^2 \rangle}$ in addition to $|M|$, which was not requested by the research specification. This completion bias reflects the tendency of LLMs to generate coherent narratives matching expected patterns regardless of whether the computation supports them. The overlapping curves obscure interpretation as visual inspection of the output figure was not performed.
Figure~\ref{fig:cases}(h) illustrates an implementation error. The agent failed to expand the bond dimension before TDVP time evolution, resulting in dynamics computed with bond dimension 1. Nearly the entire phase diagram is marked as ``Unknown/Error,'' yet the outputs appear complete and it is difficult to identify that the results are unreliable without close inspection of the underlying data.
Figure~\ref{fig:cases}(i) shows another implementation error specific to the retinal model. The highly oscillatory dynamics arise from two bugs: failing to expand the bond dimension as in Fig.~\ref{fig:cases}(h), and incorrectly handling coordinate transformations between dimensionless and dimensional coordinates. Other commonly encountered bugs for the retinal model include implementing the $\hat{p}^2$ operator as the second-order differencing operator while omitting the negative sign, and using energy values in eV directly from the literature without converting to atomic units.

The failure modes during the simulations are classified into four categories: implementation error, hallucinations, response error, and figure defects. Implementation errors occur when code runs without exceptions but produces incorrect results. These errors can arise during initial code generation, from incorrect physical parameters such as insufficient bond dimension and inadequate phonon basis truncation, or during post-processing when aggregating and analyzing results. Hallucinations encompass cases where the agent ignores user instructions, declares tasks complete prematurely, or makes judgments and decisions that contradict physical principles. Response errors arise from malformed outputs. DeepSeek-V3.2 occasionally produces responses that cannot be parsed as valid JSON, while Claude Opus 4.5 sometimes generates empty tool calls with missing arguments, both causing task termination. These response errors may stem from the complexity of maintaining structured output over extended multi-step reasoning. Finally, figure defects include visualization problems such as overlapping curves or missing labels that impair interpretation without affecting the underlying computation. We do not show results from runs terminated by response errors in Fig.~\ref{fig:cases}, as these typically produce no usable output.
We have confirmed through appropriate retry logic that such invalid responses do not arise from network errors, API rate limits, or transient failures in the agent architecture. While better prompt engineering or technical adjustments might reduce response errors, the fact that such errors occur more frequently in complex tasks like the retinal model than in simpler tasks like the Ising model suggests underlying robustness limitations in current LLMs.

\subsection{Typical Workflow}

\begin{figure}[htbp]
\centering
\includegraphics[width=\textwidth]{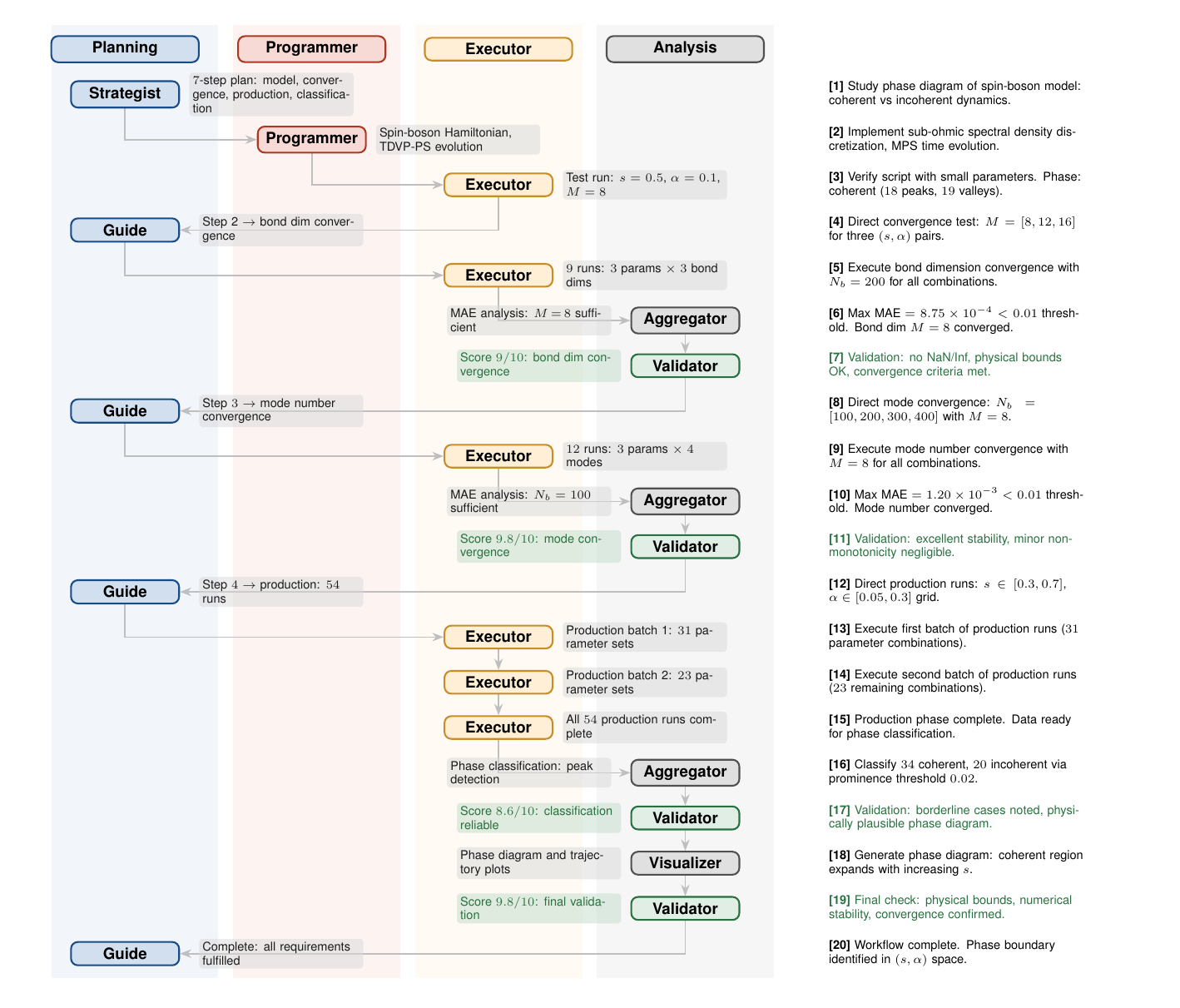}
\caption{\textbf{Complete workflow for spin-boson model phase diagram construction.} The workflow shows coordination between specialized agents across four tracks: planning (\textbf{Strategist}, \textbf{Guide}), programming (\textbf{Programmer}), execution (\textbf{Executor}), and analysis (\textbf{Aggregator}, \textbf{Validator}, \textbf{Visualizer}). The right column provides a step-by-step narrative of the 20 actions taken during the simulation. The \textbf{Validator} provides quality scores at each stage, ensuring numerical reliability before production runs and flagging borderline cases in the final phase classification.}
\label{fig:workflow}
\end{figure}

Figure~\ref{fig:workflow} illustrates a complete workflow for the sub-Ohmic spin-boson model phase diagram task completed by DeepSeek-V3.2, demonstrating how multi-agent collaboration addresses the core challenges of tensor network simulation.

Tensor network simulations require careful navigation of multiple interdependent numerical parameters. For the spin-boson model, reliable results depend on adequate bond dimension $M$ to capture entanglement, sufficient bath modes $N_b$ to represent the continuous spectral density, and appropriate time evolution settings. These parameters interact in complex ways. Insufficient bond dimension produces qualitatively wrong dynamics, while excessive values waste computational resources. Determining appropriate parameters typically requires expert judgment informed by convergence analysis, a process that is tedious, error-prone, and rarely documented in published work.

The multi-agent architecture transforms this expert-driven process into a systematic, reproducible workflow. The \textbf{Strategist} decomposes the research goal into a seven-step plan that explicitly includes convergence testing before production runs. After an initial test run verifies the implementation, the \textbf{Executor} performs convergence sweeps across bond dimensions ($M = 8, 12, 16$) and bath modes ($N_b = 100, 200, 300, 400$) for representative parameter combinations. The \textbf{Validator} confirms convergence at each stage before the \textbf{Guide} proceeds to production runs, which scan the full parameter grid with 54 simulations. The workflow concludes with phase classification and visualization, producing a publication-ready phase diagram in 20 coordinated steps.

\subsection{Performance Analysis}

We evaluate each benchmark run on a 0--10 scale (see Methods Section~\ref{sec:rubrics} for detailed rubrics). A score of 10 indicates a perfect run producing publication-quality figures with correct scientific content, while a score of 0 represents complete failure due to response errors or severe hallucinations. The successful runs shown in Fig.~\ref{fig:cases}(d--f) all receive scores of 10, demonstrating that the multi-agent architecture can produce publication-ready results. The failed runs in Fig.~\ref{fig:cases}(g--i) receive scores of 6, 4, and 3, respectively. The Ising model run (g) loses points for hallucinations and figure defects. The spin-boson run (h) and the retinal run (i) receive lower scores due to implementation errors.

\begin{figure}[htbp]
\centering
\includegraphics[width=\textwidth]{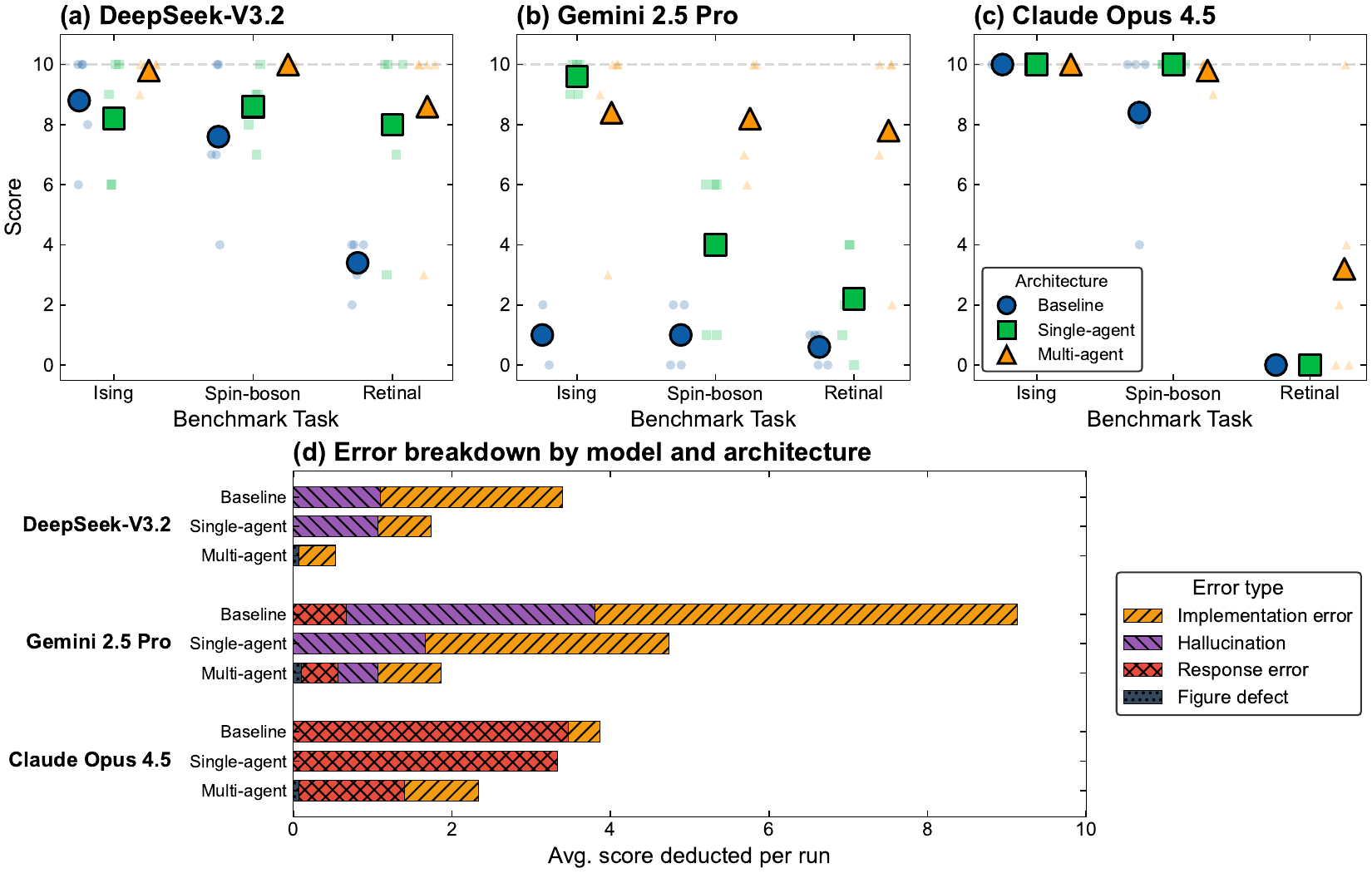}
\caption{\textbf{Performance comparison across agent architectures and LLM backends.} \textbf{(a--c)} Score distributions for each model across the three benchmark tasks (Ising, Spin-boson, Retinal). Small markers show individual runs (5 per task), while large markers indicate mean scores. \textbf{(d)} Average score deduction per run decomposed by error type for each model and architecture. Implementation errors and hallucination dominate failures for DeepSeek-V3.2 and Gemini 2.5 Pro, while response errors are the primary failure mode for Claude Opus 4.5.}
\label{fig:comparison}
\end{figure}

Figure~\ref{fig:comparison} presents a comprehensive comparison of agent performance across all three benchmark tasks and agent architectures. The top row shows score distributions for each model, revealing that multi-agent architecture consistently outperforms both baseline and single-agent configurations across all LLM backends. The baseline configuration is plagued by implementation errors, hallucinations, and response errors, with each model exhibiting distinct failure patterns. DeepSeek-V3.2 performs relatively well on the Ising and spin-boson tasks in baseline mode, but all baseline runs on the retinal model suffer from implementation errors that produce incorrect results. Gemini 2.5 Pro exhibits severe hallucinations in baseline mode across all three tasks, as the model concludes that numerical results have been obtained and figures have been plotted, when in fact the simulation code failed and nothing was produced. Claude Opus 4.5 shows promise for the Ising and spin-boson tasks in baseline and single-agent configurations, but struggles with the retinal model due to response errors. Specifically, the model generates invalid tool calls with empty arguments when attempting to write Python code files, causing immediate task termination.

The multi-agent architecture with in-context learning improves performance across all models. DeepSeek-V3.2 with multi-agent architecture achieves the highest success rate, with 87\% of runs receiving perfect scores and an average score of 9.47. Gemini 2.5 Pro shows the most dramatic improvement from multi-agent coordination. 
The baseline configuration achieves only 0.87 average score with no perfect runs, while the multi-agent configuration reaches an average score of 8.13 with 53\% of runs receiving perfect scores. 

Figure~\ref{fig:comparison}(d) decomposes the average score deduction per run by error type, revealing distinct failure patterns. For DeepSeek-V3.2 and Gemini 2.5 Pro, implementation errors and hallucinations constitute the dominant failure modes, with Gemini 2.5 Pro showing particularly severe issues in the baseline configuration.
The multi-agent architecture substantially mitigates these errors, reducing average deductions to 0.53 for DeepSeek-V3.2 and 1.87 for Gemini 2.5 Pro. Claude Opus 4.5 exhibits a qualitatively different failure pattern, where response errors account for the majority of score deductions across all architectures, primarily arising from invalid outputs in the retinal photoisomerization task. These results demonstrate that multi-agent coordination with in-context learning provides consistent improvements over baseline approaches, and that different LLM backends exhibit characteristic failure modes.

\begin{figure}[htbp]
\centering
\includegraphics[width=\textwidth]{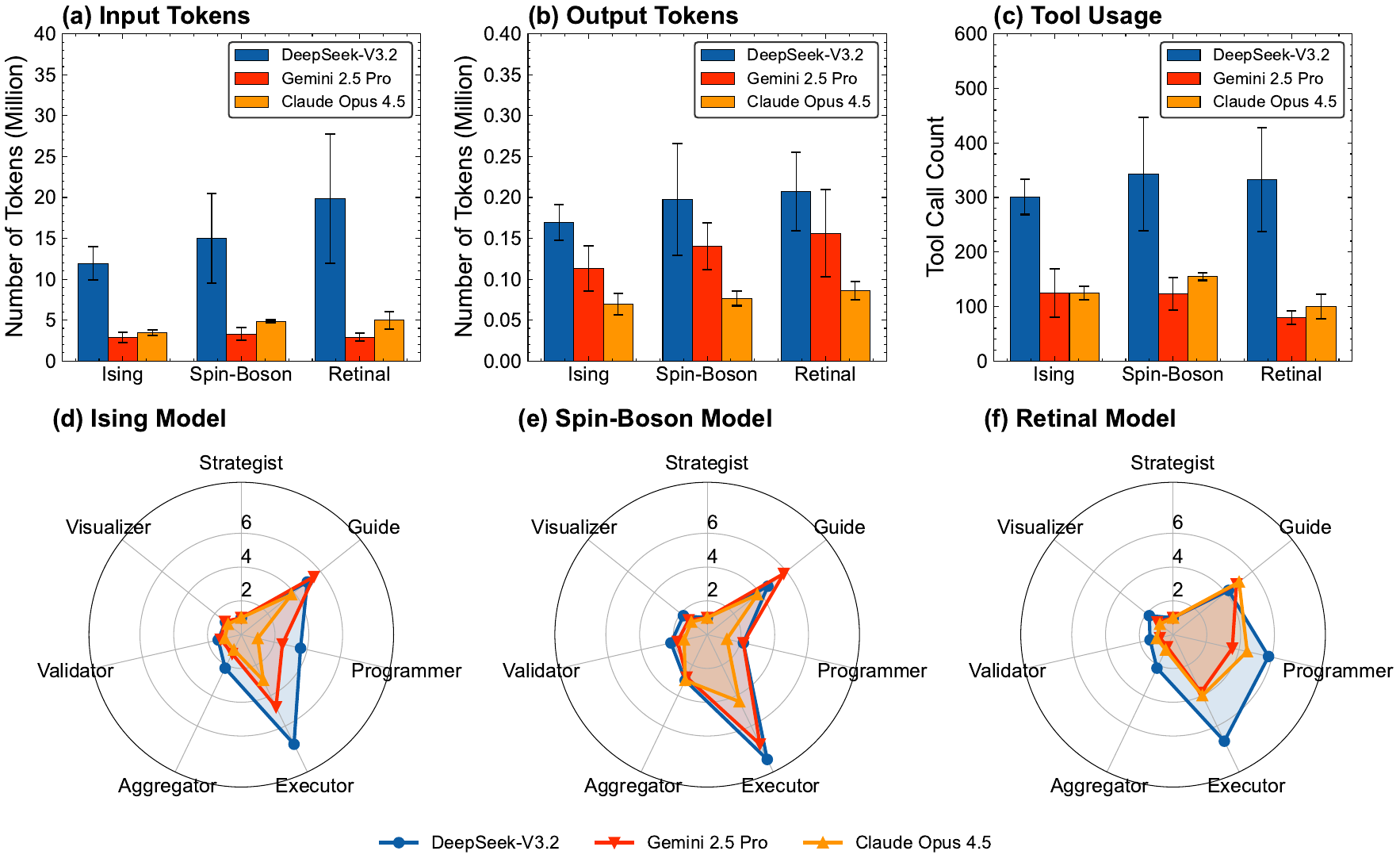}
\caption{\textbf{Token consumption and agent behavior analysis.} \textbf{(a)} Input and \textbf{(b)} output token counts for each model-task combination. \textbf{(c)} Total tool call counts. \textbf{(d--f)} Distribution of high-level agent calls across the seven specialized agents for the Ising model, spin-boson model, and retinal model, respectively.}
\label{fig:tokens}
\end{figure}

We further compare token consumption and agent behavior across tasks in Figure~\ref{fig:tokens}. Figure~\ref{fig:tokens}(a) shows input token counts for each model-task combination. The number of input tokens is similar across the three tasks for each model, indicating that all three have comparable complexity in terms of the number of steps required for completion. This similarity results from our design choice that all three tasks should finish within a reasonable time. For complex tasks such as the retinal model, we did not ask the agent to perform extensive parameter convergence tests.
Figure~\ref{fig:tokens}(b) shows output token counts, which are dramatically smaller than input tokens. Input tokens are on the order of 10 million, while output tokens are on the order of 0.1 million. This disparity arises because tool calls typically require few output tokens, whereas processing the Renormalizer documentation, script output, and full research context consumes substantial input tokens.
Across the three models, DeepSeek-V3.2 consumes significantly more tokens than the others. For example, for the Ising model, DeepSeek-V3.2 consumes roughly 12 million input tokens, while Gemini 2.5 Pro and Claude Opus 4.5 consume roughly 3--4 million tokens. Figure~\ref{fig:tokens}(c) reveals the cause of the more token consumption by DeepSeek-V3.2, which makes approximately 300 tool calls per task, compared to 100--150 for the other models.
This difference arises because DeepSeek-V3.2 tends to perform more granular actions, executing file operations such as listing directories, checking file existence, and reading files as separate calls, whereas Gemini 2.5 Pro and Claude Opus 4.5 batch operations more efficiently. For example, DeepSeek-V3.2 frequently reads simulation log files to monitor progress and check results, whereas Gemini 2.5 Pro and Claude Opus 4.5 tend to skip these intermediate checks. Since each tool call requires sending the full conversation context, more frequent calls lead to proportionally higher token consumption.
Despite higher token consumption, DeepSeek-V3.2 remains cost-effective due to its lower API price. During the course of this study, the API prices for the three models were approximately \$0.3, \$1.25, and \$5 per million tokens, respectively. This makes DeepSeek-V3.2 the most cost-effective option among the three. Gemini 2.5 Pro has a slightly higher cost, while Claude Opus 4.5 is typically 5 to 10 times more expensive than the other two models. The typical cost per task is estimated to be \$2--3 for DeepSeek-V3.2 and Gemini 2.5 Pro, and \$20--30 for Claude Opus 4.5. Cache reading may further reduce costs but is not considered here.

The radar plots in Figure~\ref{fig:tokens}(d--f) show the distribution of high-level agent calls to the seven specialized agents for each model-task combination.
The Ising and spin-boson model tasks require extensive numerical calculations, resulting in more calls to the \textbf{Executor} and \textbf{Aggregator}. In contrast, the major challenge for the retinal model is correct implementation of the simulation code, leading to relatively more calls to the \textbf{Programmer}.
Among the LLM models, DeepSeek-V3.2 exhibits a higher number of agent calls, reflecting more iterative debugging cycles and data processing steps.

During preliminary development, we tested DeepSeek-V3.2 without reasoning enabled and observed severe hallucination in the \textbf{Validator}. For example, the \textbf{Validator} frequently claimed that Ising model results were incorrect because ``the energy should increase with $h$,'' and suggested the Hamiltonian implementation was wrong.When tested in a standalone chat environment without reasoning, DeepSeek-V3.2 almost consistently answered that energy increases with transverse field when asked to respond quickly without explanation. 
After receiving the warning from the \textbf{Validator}, the \textbf{Conductor} would then attempt to verify the result, and many steps were taken before realizing the Validator had made a mistake.
This hallucination disappeared when reasoning was enabled, motivating our decision to use reasoning-enhanced models for all production runs reported in this work.
This suggests that scientific agents must be designed to encourage deliberate reasoning rather than rapid response, particularly for validation tasks where physical intuition matters.

\section{Discussion}

We have demonstrated that LLM-driven agents can autonomously perform tensor network simulations of quantum many-body systems. Our approach addresses two key challenges: the sparse coverage of tensor network methods in LLM training data through in-context learning with curated documentation, and the tight coupling between physics, code, and data through multi-agent decomposition.
Systematic comparison across three agent configurations reveals that both components are essential. The baseline configuration, which relies on the LLM to search through source code, fails frequently due to hallucinations and implementation errors. Adding in-context learning substantially improves performance, while multi-agent decomposition provides further gains by enabling focused expertise and iterative validation. The multi-agent architecture achieves approximately 90\% success rate across three representative tasks spanning ground-state calculations, open quantum system dynamics, and photochemical reactions.

Analysis of both successful and failed runs reveals characteristic patterns across LLM backends. DeepSeek-V3.2 achieves the highest success rate but consumes more tokens due to more granular actions. Gemini 2.5 Pro shows the most dramatic improvement from multi-agent coordination, rising from near-complete failure in baseline mode to reasonable performance with the full architecture. Claude Opus 4.5 achieves consistent results but occasionally produces malformed outputs that terminate workflows prematurely.

Despite these encouraging results, the generated outputs require human verification due to the non-negligible error rate. In practice, failures are typically detectable, because the \textbf{Validator} flags convergence issues, response errors terminate workflows with clear error messages, and implementation errors often produce obviously unphysical results (as in Fig.~\ref{fig:cases}(i)).
Some failures, however, produce plausible-looking but incorrect results that require expert inspection to identify, representing a minority of cases. The failure modes we observed, including numerical parameter insufficiency, implementation errors despite correct reasoning, and subtle mistakes in translating mathematical specifications to code, mirror challenges faced by human researchers, suggesting that complete automation of scientific computation remains an open problem.

Looking forward, we anticipate rapid progress on several fronts. Continued improvements in LLM reasoning capabilities and context windows will likely reduce error rates and expand the complexity of problems that can be tackled autonomously. Advances in agent architectures, including improved multi-agent collaboration and self-verification protocols, may enhance output quality. Integration with literature retrieval could enable agents to autonomously identify research questions and design simulations. The emergence of multi-modal models will enable agents to directly parse equations, tables, and figures from published literature, and to critically examine their own generated plots for anomalies, iteratively refining them.

As these capabilities mature, LLM-driven agents may democratize access to tensor network methods. Currently, effective use of these powerful techniques requires years of specialized training, limiting their application to a small community of experts. By encoding domain expertise in documentation and agent workflows, the barrier to entry could be dramatically lowered, enabling researchers from adjacent fields to leverage tensor network methods for their own problems. Domain experts, freed from routine implementation and convergence test tasks, could focus on more creative and challenging aspects of research, accelerating scientific discovery in quantum many-body physics and beyond. 

\section{Methods}

\subsection{Matrix Product States and Renormalizer}
The matrix product state is a representative tensor network ansatz that decomposes the exponentially large state vector into a product of local tensors. For a system of $N$ sites, an MPS represents the state as
\begin{equation}
\label{eq:mps}
    \ket{\Psi} = \sum_{\{a\},\{\sigma\}}
     A^{\sigma_1}_{a_1} A^{\sigma_2}_{a_1 a_2} \cdots
           A^{\sigma_N}_{a_{N-1}} \ket{\sigma_1 \sigma_2 \cdots \sigma_N}
\end{equation}
where $\sigma_i$ labels the local basis states with dimension $d$, and $a_i$ are auxiliary bond indices with dimension up to $M$, the bond dimension. The bond dimension controls the amount of entanglement that can be captured. Larger $M$ allows more accurate representation of highly entangled states but increases computational cost, which scales as $O(NM^3d)$ for typical operations. For ground state calculations, the density matrix renormalization group algorithm optimizes the MPS tensors variationally. All results reported in this work use MPS as the simulation ansatz.

For time evolution, the time-dependent variational principle projects the Schr\"odinger equation onto the MPS manifold. The key idea is to find the optimal time derivative of the MPS that best approximates the exact evolution within the manifold of fixed bond dimension. This is achieved by applying the projector $\hat{P}_{T_{|\Psi\rangle}\mathcal{M}}$ onto the tangent space of the MPS manifold at the current state:
\begin{equation}
\label{eq:tdvp}
    \frac{\partial}{\partial t} \ket{\Psi} = -i \hat{P}_{T_{|\Psi\rangle}\mathcal{M}} \hat{H} \ket{\Psi}
\end{equation}
The projector can be decomposed into local terms involving forward and backward sweeps through the MPS sites, leading to an efficient algorithm with computational cost scaling as $O(NM^3d)$ per time step. TDVP preserves important physical properties including unitarity and energy conservation, making it particularly suitable for long-time dynamics simulations.

Renormalizer is an open-source Python package for tensor network simulations with a special focus on electron-phonon quantum dynamics~\cite{renormalizer, ren2018time}. The package provides MPS and matrix product operator algorithms for ground state search and time evolution, tree tensor network support for more complex network topologies, and automatic operator construction from first and second-quantized Hamiltonians~\cite{ren2020general, li2024optimal}. Renormalizer also supports finite temperature simulations through imaginary time evolution or thermofield transformation, GPU acceleration via CuPy, and quantum number conservation for exploiting symmetries. A comprehensive review of the underlying methods is available in Ref.~\cite{ren2022time}.

\subsection{Agent Implementation and LLM Configuration}

The agent system is implemented using the LangChain framework for LLM orchestration. Each agent is instantiated as a LangChain tool with a dedicated system prompt that encodes its specialized role and constraints.
The \textbf{Conductor} operates with a recursion limit of 200 steps, allowing complex multi-step workflows while preventing infinite loops. All agents share access to common utility tools for file operations, including reading, writing, and listing directories, as well as code execution.
The \textbf{Strategist} and \textbf{Guide} roles are implemented as a single LLM with shared conversation history, enabling coherent planning across multiple consultation rounds. A call to the \textbf{Strategist} generates a detailed step-by-step research plan, while subsequent calls to the \textbf{Guide} continue the same conversation thread, allowing the \textbf{Guide} to track progress against the original plan.
The \textbf{Programmer} agent receives requirements from the \textbf{Conductor} along with the original user input for context and uses a reasoning-enhanced model at temperature 0 to ensure careful code generation.
The \textbf{Executor} agent runs simulations by invoking the main script with different parameter sets. It supports parallel execution of up to 32 jobs per batch and automatically handles job scheduling for larger parameter sweeps. When encountering errors, the \textbf{Executor} attempts to debug by analyzing log files and applying fixes via diff patches.
The \textbf{Aggregator} agent processes output files from multiple simulation runs, computing derived quantities and preparing data for visualization and human inspection. It answers specific scientific questions posed by the \textbf{Conductor} using quantitative analysis of the simulation outputs.
The \textbf{Validator} agent critically examines numerical results for unphysical values, convergence failures, and systematic errors, providing severity assessments with recommendations to continue, investigate, or stop execution.
The \textbf{Visualizer} agent generates publication-quality figures with a custom style template. It inspects data structures, makes plots using matplotlib, and saves output as PDF files.

We evaluate three LLM backends: DeepSeek-V3.2 with reasoning enabled, Gemini 2.5 Pro, and Claude Opus 4.5. All models use temperature 0 for the \textbf{Conductor} and \textbf{Programmer} agents to ensure consistent outputs, while other agents use default temperature settings. The system implements automatic retry logic with exponential backoff (initial delay of 120 seconds, factor of 2.0, maximum of 3 retries) to handle response errors. Invalid tool calls trigger automatic retries with the same backoff policy.

\subsection{Benchmark Tasks}

The three benchmark tasks were selected according to the following criteria: (1) coverage of distinct algorithmic challenges (ground-state optimization, real-time dynamics, and custom basis implementation), (2) representation of different physical domains where tensor networks are applied (condensed matter, open quantum systems, and photochemistry), (3) varying degrees of information availability in LLM training data (well-documented critical point versus debated phase boundary versus specialized model), and (4) completion within a reasonable time frame (hours rather than days) to enable systematic evaluation across multiple runs and configurations.

\subsubsection{Two-Dimensional Transverse-Field Ising Model}

The two-dimensional transverse-field Ising model on an $N_x \times N_y$ lattice is described by the Hamiltonian
\begin{equation}
\label{eq:ising}
    \hat{H} = -J \sum_{\langle i,j \rangle} \sigma_i^z \sigma_j^z - h \sum_i \sigma_i^x
\end{equation}
where $J > 0$ is the ferromagnetic coupling between nearest-neighbor spins, $h$ is the transverse field strength, and $\sigma_i^{x,z}$ are Pauli matrices at site $i$. The ground state of this model is governed by the competition between $J$ and $h$. At zero temperature, the system undergoes a quantum phase transition as the ratio $h/J$ is varied. For small $h/J$, the ground state exhibits spontaneous magnetization in the ferromagnetic phase, while for large $h/J$, quantum fluctuations dominate and the system becomes paramagnetic.
The quantum transverse-field Ising model in two dimensions has no exact analytical solution, and extensive numerical studies using cluster Monte Carlo and DMRG have established the critical field at $h_c/J \approx 3.04$ in the thermodynamic limit~\cite{de1998critical, blote2002cluster}. The order parameter $|M_z|$ versus $h/J$ provides a direct probe of the phase transition.

We tasked our LLM-driven agent with studying the quantum phase transition in this model. We limited the computation to an $8 \times 8$ lattice. Although finite-size effects are substantial at this system size, this choice ensures that calculations complete within a reasonable time frame. The methodology readily generalizes to larger lattices for production research.

\subsubsection{Sub-Ohmic Spin-Boson Model}

The spin-boson model describes a two-level system coupled to a bosonic bath and serves as a paradigmatic model for quantum dissipation~\cite{leggett1987dynamics}. The Hamiltonian is
\begin{equation}
\label{eq:sbm}
    \hat{H} = \frac{\Delta}{2} \sigma_x + \frac{\epsilon}{2} \sigma_z + \sum_k \omega_k b_k^\dagger b_k + \frac{\sigma_z}{2} \sum_k g_k (b_k^\dagger + b_k)
\end{equation}
where $\Delta$ is the tunneling amplitude, $\epsilon$ is the bias, $\omega_k$ and $g_k$ are the frequency and coupling strength of the $k$-th bath mode, and $b_k^{(\dagger)}$ are bosonic annihilation (creation) operators. The bath is characterized by the spectral density
\begin{equation}
\label{eq:spectral}
    J(\omega) = 2\pi\alpha \omega_c^{1-s} \omega^s e^{-\omega/\omega_c}
\end{equation}
where $\alpha$ is the dimensionless coupling strength, $\omega_c$ is the cutoff frequency, and $s$ determines the bath type: sub-Ohmic ($s < 1$), Ohmic ($s = 1$), or super-Ohmic ($s > 1$).
The sub-Ohmic spin-boson model exhibits a quantum phase transition at zero temperature. Below a critical coupling $\alpha_c(s)$, the spin dynamics shows coherent oscillations, while above $\alpha_c$, the dynamics becomes incoherent with monotonic decay toward a localized state. The boundary between coherent and incoherent regimes defines a dynamical phase diagram in the $\alpha$-$s$ parameter space.
Determining this phase boundary has been the subject of extensive numerical investigation using a variety of advanced methods, including multilayer multiconfiguration time-dependent Hartree~\cite{wang2010coherent}, the Davydov ansatz~\cite{wang2016variational}, hierarchical equations of motion~\cite{duan2017zero}, and quasi-adiabatic propagator path integral~\cite{otterpohl2022hidden}. Remarkably, despite decades of effort, the exact location of the coherent-incoherent phase boundary remains under debate. 

For numerical simulations, the continuous bath is discretized into $N_b$ modes using an exponential discretization scheme. The mode frequencies are determined by
\begin{equation}
\label{eq:discretization}
    \omega_k = -\omega_c \ln\left[1 - \frac{k}{N_b+1}\right], \quad k = 1, 2, \ldots, N_b
\end{equation}
with corresponding coupling strengths derived from the spectral density. The exponential discretization scheme in Eq.~\ref{eq:discretization} implicitly determines the discrete frequencies $\omega_k$ by inverting a cumulative distribution. Deriving this expression requires correctly evaluating the integral and obtaining the analytical form $\omega_k = -\omega_c \ln[1 - k/(N_b+1)]$. All three LLM backends successfully performed this derivation.

Obtaining converged results requires careful selection of numerical parameters. The agent must perform convergence analysis over the bond dimension $M$ and the number of bath modes $N_b$, and choose an appropriate basis truncation for each bosonic mode to ensure reliable dynamics.

\subsubsection{Retinal Photoisomerization Model}

The retinal photoisomerization in rhodopsin represents one of the fastest and most efficient photochemical reactions in nature. Upon absorption of a photon, the 11-cis retinal chromophore undergoes isomerization to the 11-trans configuration. This ultrafast reaction has attracted widespread interest due to its importance as the first step in vision~\cite{hahn2000femtosecond, lyu2022tensor, liu2024benchmarking}.
We employ a well-established two-state 25-mode model to describe the retinal photoisomerization reaction. The model comprises two diabatic electronic states $|0\rangle$, the ground state S$_0$, and $|1\rangle$, the excited state S$_1$, one torsional degree of freedom $\theta$ describing rotation about the C$_{11}$=C$_{12}$ double bond, one stretching mode $R_c$ that mediates electronic transitions near the conical intersection, and a bath of 23 harmonic oscillators $\{R_j\}$ representing the protein environment. The full Hamiltonian takes the form:
\begin{equation}
\hat{H} = \hat{H}_0 |0\rangle\langle 0| + \hat{H}_1 |1\rangle\langle 1| + \hat{V}_{01}(|0\rangle\langle 1| + |1\rangle\langle 0|)
\label{eq:retinal}
\end{equation}
where the diabatic state Hamiltonians are:
\begin{align}
\hat{H}_0 &= \frac{\hat{p}_\theta^2}{2I} + \frac{W_0}{2}(1 - \cos\theta) + \hat{H}_{\text{vib}} \\
\hat{H}_1 &= \frac{\hat{p}_\theta^2}{2I} + E_1 - \frac{W_1}{2}(1 - \cos\theta) + \hat{H}_{\text{vib}} + \kappa_c \hat{R}_c + \sum_{j=3}^{25} \kappa_j \hat{R}_j
\end{align}
Here $I$ is the moment of inertia for torsional motion, $W_0$ and $W_1$ are the barrier heights for the ground and excited state potentials respectively, $E_1$ is the vertical excitation energy, and the $\kappa$ terms represent linear vibronic coupling. The vibrational Hamiltonian $\hat{H}_{\text{vib}}$ describes the harmonic bath modes:
\begin{equation}
\hat{H}_{\text{vib}} = \frac{\Omega_c}{2}(\hat{p}_c^2 + \hat{R}_c^2) + \sum_{j=3}^{25} \frac{\omega_j}{2}(\hat{p}_j^2 + \hat{R}_j^2)
\end{equation}
The diabatic coupling is mediated by the coupling mode, $\hat{V}_{01} = \lambda \hat{R}_c$, where $\lambda$ controls the strength of non-adiabatic transitions.

This benchmark poses a distinctive challenge of implementing a specialized basis set not available in Renormalizer. The torsional coordinate $\theta$ is periodic with period $2\pi$, requiring basis functions that respect this periodicity. The appropriate choice is the exponential discrete variable representation (exponential DVR).
Another challenge is the use of dimensionless coordinates $\hat R$ and momenta $\hat p$ in the Hamiltonian, a convention that differs from both the spin-boson model described above and the default convention in Renormalizer. The agent must correctly identify this difference and implement the Hamiltonian accordingly. This knowledge is typically acquired through graduate-level training in quantum dynamics.

\subsection{Prompts of the Benchmark Tasks}
For the Ising model task, the prompt contains minimal information, as described in the Results section.
For the spin-boson model task, the prompt includes the necessary equations for deterministic implementation, namely the Hamiltonian and the exponential discretization of the spectral density.
The prompt also includes an explicit definition of the phase classification criterion to ensure reproducible results across different runs. Coherent dynamics is defined as damped oscillatory behavior exhibiting at least one valley and one subsequent peak in $\langle \sigma_z(t) \rangle$, while incoherent dynamics encompasses either purely monotonic decay or a single valley followed by localization without oscillation. This level of methodological detail is necessary because the coherent-incoherent boundary is precisely where dynamics are most ambiguous.
For the retinal model task, the prompt provided to the agent consists of a description of the model from the literature containing the complete model specification, including all Hamiltonian parameters and bath mode frequencies~\cite{liu2024benchmarking}. The description was generated by taking a screenshot of the published paper and converting it to LaTeX using LLMs. Similarly, the agent was provided with a description of exponential DVR from the literature~\cite{beck2000multiconfiguration}. The agent was instructed to run simulations with two different bond dimensions, 70 and 128.
Details of all prompts are provided in the Supporting Information.

\subsection{Evaluation Rubrics}
\label{sec:rubrics}
Each benchmark run is evaluated on a 0--10 scale based on the quality of the simulation code, numerical results, and final interpretation and visualization. The rubrics are designed to capture the error types discussed in the Results section, including implementation errors, hallucinations, response errors, and figure defects.

A score of 10 represents a perfect run, which means that the agent produces publication-quality figures with correct scientific content and professional interpretation, fully addressing the research specification. A score of 8 indicates near-perfect execution where the results are scientifically correct but minor issues exist, such as suboptimal figure aesthetics or incomplete adherence to the research specification that does not affect scientific conclusions. This score typically results from figure defects or minor implementation errors during post-processing and analysis.

A score of 6 reflects runs where the simulation code executes correctly and produces some valid results, but significant errors occur during execution or analysis. This includes cases where key information required by the research specification is missing from the produced figure, or where some scientific content in the figure is incorrect. Such scores typically arise from major implementation errors or hallucinations during the analysis phase. A score of 4 indicates partial failure, which means that the code is nearly correct but contains incorrect parameters or 1--2 bugs that an expert could easily fix, with only partial results of limited reference value. This score is typically caused by major implementation errors in the main simulation script or strong hallucinations during execution and analysis.

A score of 2 represents mostly failed runs where code implementation is incomplete with significant bugs, producing no usable results. A score of 0 indicates complete failure: no working code is produced, the agent terminates due to response errors, or severe hallucinations result in entirely unreliable outputs. Scores of 0--2 are typically associated with severe hallucinations, fundamental implementation errors, or response errors that terminate the workflow prematurely.

\section*{Data Availability}

The Renormalizer package is available at \url{https://github.com/shuaigroup/Renormalizer}. The source code of the agent system and the log files for all benchmark tasks will be made available upon publication.

\section*{Acknowledgements}
The authors thank Shi-Xin Zhang, Xiang Sun, Zhenggang Lan, and Ningyi Lyu for insightful discussions.
Weitang Li is supported by the Guangdong Basic Research Center of Excellence for Aggregate Science, the Shenzhen Science and Technology Program (No. KQTD20240729102028011), and the National Natural Science Foundation of China (Grant No. 22595405).
Jiajun Ren is supported by the National Natural Science Foundation of China (Grant No. 22273005 and No. 22422301).

\section*{Competing Interests}

The authors declare no competing interests.

\bibliography{refs}

\end{document}